# Optimal Feedback Systems with Analogue Adaptive Transmitters
Anatoliy Platonov, *Senior Member, IEEE*
*Abstract*–The paper presents original approach to concurrent optimization of the transmitting and receiving parts of adaptive communication systems (CS) with feedback channels. The results of research show a possibility and the way of designing the systems transmitting the signals with a bit rate equal to the capacity of the forward channel under given bit-error rate (BER). The results of work can be used for design of different classes of high-efficient low energy/size/cost CS, as well as allow further development and extension.

*Index Terms*— Feedback systems, analogue transmission, full optimisation, power/spectral efficiency, information limits.

## I. Introduction

The main task of information and communication theory is approaching the power/bandwidth efficiency of CS to the Shannon's boundary [1],[2], (see Fig.1, also Sect. 5). However, strict analytical results enabling systematic design of CS which work at this boundary have not been obtained.

In the paper, we show a possibility to solve this problem for a special class of feedback CS (FCS). The obtained results determine conditions sufficient for designing the system whose power/bandwidth efficiency attains the Shannon's boundary. Particularity of the system is analogue transmitting unit (TU) which is realized as adaptive pulse-amplitude (PAM) modulator adjusted by the controls computed in receiving base station (BS) and delivered to TU through the feedback channel (see Fig.2). Lack of coding/decoding units permits to describe both parts of the considered adaptive FCS (AFCS) in continuous variables. This permits to construct full mathematical model of the system, accessible for analysis fidelity criterion and to solve the optimization task using known methods of optimal estimation theory.

In the paper, full optimization task is considered, i.e. concurrent optimization of the transmitting and receiving parts of the AFCS. Formulation and solution of this task are based on the approach described in [3],[4]. Initial results of AFCS full optimization were presented in [5] and, in part, in [4]. The present paper extends and deepens results of these works. To clarify the analysis of new effects, simplest single input - single output (SISO) AFCS is considered.

The transmitting part of AFCS (Fig.1) consists of the sample and hold (S&H) block and adaptive modulator, which contains the subtracting unit ($\Sigma$) and PAM modulator – transmitter (M1) with adjusted modulation depth.

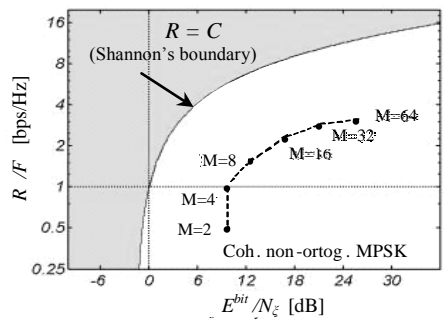

Fig.1 Illustration of power-bandwidth tradeoff in CS with coherent non-orthogonal MPSK [1].

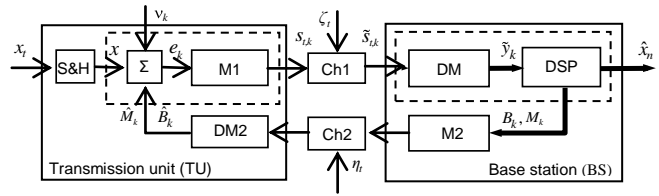

Fig. 2. Block-diagram of adaptive feedback CS (AFCS).

Base station (receiving and controlling part of the system) includes the analogue receiver – demodulator (DM1) and digital signal processing unit (DSPU). DSPU computes the estimates of input signal and controls which are transmitted to TU through the feedback channel M2-Ch2-DM and used for adjusting the modulator. There is assumed that forward (Ch1) and feedback (Ch2) channels are stationary, memoryless channels with additive white Gaussian noises (AWGN). Notations in Fig. 2 and general principles of AFCS functioning are explained in Section II.

One should say that the task of analogue FCS optimization was considered by many authors. In basic works [8]-[13] and others, the results extending Shannon's rate distortion theory [6],[7] to the analogue communication were obtained. However, further researches in the field were somewhat hampered by spectacular successes of digital communications theory and its industrial implementations. Nevertheless, interest in the analogue transmission did not disappear (e.g. [14],[15]).

From our point of view, decreased attention to the analogue FCS in last decades was caused also by impossibility to implement the theoretical results due to commonly used linear model of TU. This excludes a possibility to consider over-modulation of the transmitter which drastically disturbs the work of FCS and should be taken into account in formulation and solution of the optimization task. In [3]-[5] and in the present paper, this is done by direct introduction of non-linear model of PAM modulator (Fig.3) and "statistical fitting condition", which imposes necessary limitations on TU parameters ensuring exclusion of over-modulation at the given level of confidence. Fulfillment of fitting condition permits to consider TU as practically always (with a probability not exceeding given small value $\mu$) linear unit. In this case, optimization of AFCS can be carried out using known methods of optimal estimation theory [15],[16]. Extreme of the fidelity criterion should be searched under fitting condition as additional constraint ([3]-[5], see also Sect. III). Less but also crucial reason restricting implementation of results was assumption of noiseless feedback channel.

The paper is organized as follows. In Section II, principles of AFCS work, mathematical models of the main units and full model of the system are described. Section III is devoted to formulation and solution of (Bayesian) full optimization task. New effects appearing in the optimal AFCS and particularities of their work are analyzed in Section VI. In Section V, information characteristics and power/spectral efficiency of optimal AFCS, as well as their connections with known results are discussed. Concluding remarks are drawn in Section VI.

## II. FUNCTIONING AND FORMAL DESCRIPTION OF ACS

We assume that the signals $x_t$ at the TU input are band-limited Gaussian processes with known mean value $x_0$ and variance $\sigma_0^2$. After sampling in the sample-and-hold (S&H) unit, each sample $x^{(m)} = x(mT)$, ($m=1,2,...$) is held at the input of subtractor $\Sigma$ during the time $T = 1/2F$, ($m=1,2...$; $2F$ is two-side baseband of the signal $x_t$). Each sample $x^{(m)}$ is transmitted independently and in the same way in $n = T/\Delta t_0$ cycles ($\Delta t_0 = 1/2F_0$ is a duration of the single cycle, $F_0$ determines two-side bandwidth of forward and feedback channels). Under these conditions, analysis of AFCS can be reduced to the analysis of a single sample transmission that permits to omit upper indices in notations of the samples ($x^{(m)} = x$) and related variables.

Remark: under formulated above assumptions value

$$n = \frac{T}{\Delta t_0} = \frac{F_0}{F} \quad (1)$$

determines the coefficient of extension of the input signal spectrum, and $F$ determines the band-pass of the AFCS.

In each $k$-th cycle ($k=1,...,n$), BS processes the signal received from TU and computes intermediate estimate $\hat{x}_k$ of the sample stored it until the next cycle. Simultaneously, it computes the control signals $M_k, B_k$ transmitted to TU through the feedback channel. We assume that duration $\Delta t_0$ of the cycles and distance between TU and BS are sufficient for these signals were delivered to TU and used for a setting the parameters of its units before the beginning of the next cycle of conversion. After $n$ cycles, final estimate $\hat{x}_n$ of the sample $x$ is directed to external addressee, units of AFCS are reset to the initial state, and next sample transmission begins. To simplify the analysis, PAM modulator with double side band suppressed carrier (DSB-SC [1]) is considered. Extension of results to AFCS with full or SSB AM requires only recalculation of the power of emitted signals.

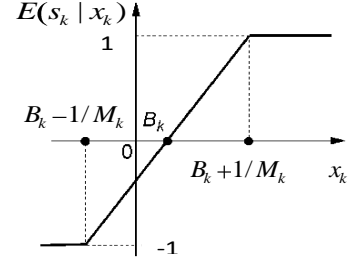

Fig. 3. Static transition function of the adaptive modulator.

In this case, mathematical model of adaptive PAM modulator and transmitter can be presented, for each $k = 1,...,n$, by the relationship:

$$s_{t,k} = A_0 \begin{cases} M_k(x_k - B_k) & \text{if } M_k|x_k - B_k| \leq 1 \\ \text{sign}(x_k - B_k) & \text{if } M_k|x_k - B_k| > 1 \end{cases} \cos(2\pi f_0 t + \varphi_0) \quad (2)$$

where $s_{t,k}$ is the emitted signal, $A_0, f_0, \varphi_0$ are amplitude, frequency and phase of the carrier signal, respectively. Unlike the conventional modulators, modulation depth $M_k$ in (2) may take different (in general case, adaptively adjusted) values in sequential cycles of the sample transmission. The second adaptively adjusted parameter $B_k$ determines the position of characteristic of modulator (see Fig. 2). Signal $x_k = x + v_k$ at the subtractor $\Sigma$ input is a sum of the input sample $x$ and zero mean Gaussian internal noise $v_k$ with the variance $\sigma_v^2$. Difference signal acting at the modulator M1 input has the form: $e_k = x_k - B_k = x - B_k + v_k$.

Model (2) allows us to analyze the work of AFCS directly taking into account possible over-modulation which causes emission of the carrier signal $s_{t,k}^{car} = A_0 \cos(2\pi f_0 t + \varphi_0)$ without any information about the sample. The result is appearance of abnormal errors in estimates $\hat{x}_k$.

Remark. According to (2), mean power of the emitted signals is always less than $A_0^2/2$ and equal to $A_0^2/2$ when the difference signal $x_k - B_k$ exceeds the levels $\pm 1/M_k$. The latter means that model (2) is equivalent to the model of TU with linear modulator M1 and emitter with overloading on the level $A_0$.

Adaptive adjusting of the parameters $M_k, B_k$ is realized using controls computed in DSPU and transmitted to TU through the feedback channel Ch2. As it is shown below, (formula (19) in Sect III.C, see also [5]), in Gaussian case, optimal values of modulation depth $M_k$ do not depend on information delivered to DSPU in previous cycles and can be determined at the initial stage of AFCS design. In practice, sequential switching the gains $M_k$ to optimal values can be realized using amplifier with controlled gain or the bank of amplifiers with properly set gains. In turn, parameters $B_k$ have continuous values and depend on the control signals computed by DSPU and transmitted to TU.

Remark: in the noiseless stationary feedback channel Ch2 the received and transmitted values of $B_k$ are the same. In the noisy feedback channels, the received in TU value $\hat{B}_k$ (estimate of $B_k$) can be presented as the sum: $\hat{B}_k = B_k + \Delta B_k$. In linear feedback channels with AWGN $\eta_k$, transmission errors $\Delta B_k$ are also Gaussian zero-mean random values with the variance $\sigma^2_{B,k} = \sigma^2_B$. This permits, without loss of generality, to include the feedback noise $\Delta B_k$ as (in many cases, dominating) component, into internal noise $v_k$ increasing its variance by $\sigma^2_B$. Such a transition allows us to analyze AFCS with noisy channel Ch2, as if it was the noiseless channel and assume $\hat{B}_k = B_k$.

To extend possibilities of the analysis, below we take into account influence of the distance $r$ between TU and BS. In this case, model of the signal $\tilde{s}_{t,k}$ at the input of receiver DM1 takes the form:

$$\tilde{s}_{t,k} = \frac{\gamma}{r} s_{t,k} + \zeta_t \quad (3)$$

where $\zeta_t$ is AWGN with the variance $\sigma^2_\zeta$, and $\gamma$ is the gain of the channel Ch1. It is assumed that the distance between TU and BS, as well as the gain $\gamma$ remain constant during the time $T$ of the sample transmission.

After demodulation and digitizing of the received signal $\tilde{s}_{t,k}$ in block DM1 of BS, signal ("observation") routed to the input of DSPU takes the form

$$\tilde{y}_k = A \begin{cases} M_k(x_k - B_k) & \text{if } M_k |x_k - B_k| \leq 1 \\ \text{sign}(x_k - B_k) & \text{if } M_k |x_k - B_k| > 1 \end{cases} + \xi_k \quad (4)$$

where $A = A_0 \gamma / r$, and $\xi_k$ is AWGN with the variance $\sigma^2_\xi = F_0 N_\xi$, where $N_\xi / 2$ is double-side spectral power density of the noise in forward channel Ch1.

One should notice that mean power of the emitted by TU signal and signal-to-noise ratio (SNR) at the channel Ch1 output depend on the type of AM. These differences can be taken into account in the value of the gain $\gamma$ and further (4) is used as a model of the channel M1-Ch1-DM1 independently from the type of modulation. Quantization noise of A/D converter at the DM1 output is assumed to be negligibly small and is not considered.

Block DSPU (digital receiver of BS) computes intermediate estimates $\hat{x}_k = \hat{x}_k(\tilde{y}_1^k)$ of the transmitted sample according to the Kalman-type equation [16]:

$$\hat{x}_k = \hat{x}_{k-1} + L_k[\tilde{y}_k - E(\tilde{y}_k | \tilde{y}_1^{k-1})] \quad (5)$$

where gains $L_k$ determine the rate of estimates convergence, $E(\tilde{y}_k | \tilde{y}_1^{k-1})$ describe the predicted values of observations $\tilde{y}_k$, and $\tilde{y}_1^{k-1}$ denotes the sequence of observations: $\tilde{y}_1^{k-1} \triangleq (\tilde{y}_1, ..., \tilde{y}_{k-1})$. Initial condition for (5): $\hat{x}_0 = x_0$, where $x_0$ is mean value of the input signal. Simultaneously, for each $k = 1, ..., n$, DSPU computes controls $B_k = B_k(\tilde{y}_1^{k-1})$ transmitted to TU via the feedback channel Ch2.

Summary: AFCS transmits the sample and forms its final estimate $\hat{x}_n$ in $n$ cycles (iterations) independently from the previous samples and estimates. In each cycle $k = 1, ..., n$, adaptive modulator $\Sigma$+M1 forms the signal $s_{t,k} = A_0 M_k (x_k - B_k) \cos(2\pi f_0 t + \varphi_0)$ emitted into channel Ch1. The received signal $\tilde{s}_{t,k}$ is demodulated and digitized in demodulator DM1. The obtained code (observation) $\tilde{y}_k$ is routed to the digital unit DSPU which computes new estimate $\hat{x}_k$ and control $B_{k+1} = B_{k+1}(\tilde{y}_1^k)$. The computed value $B_{k+1}$ is transmitted to TU, and estimate $\hat{x}_k$ is stored in DSPU until the next cycle. Receiving the value $B_{k+1}$ finishes the cycle. After reinitializing corresponding units of AFCS and setting the parameters $M_{k+1}, B_{k+1}$ and $L_{k+1}$ to the new values, $(k+1)$-th cycle of transmission begins. After $n$ cycles, final estimate $\hat{x}_n$ of the sample is routed to the external addressee, and AFCS begins transmission of the next sample.

### III. FULL OPTIMIZATION OF AFCS

Known prior distribution of the samples and models introduced in Sect. II permit to formulate the criterion of the transmission quality - mean square errors (MSE) of current estimates of the sample:

$$P_k = E[(x - \hat{x}_k)^2] = \int_{-\infty}^{\infty} ... \int_{-\infty}^{\infty} [x - \hat{x}_k(\tilde{y}_1^k)]^2 p(x/\tilde{y}_1^k) p(\tilde{y}_1^k) dx d\tilde{y}_1^k \quad (6)$$

where $d\tilde{y}_1^k = \prod_{i=1}^k d\tilde{y}_i$ posterior probability density function (PDF) of the sample values $p(x/\tilde{y}_1^k) = p(x_k/\tilde{y}_1^k, B_1^k, M_1^k)$ depends on the values of parameters $B_1^k, M_1^k$ in previous and current cycles of transmission. Explicit form of PDF $p(x/\tilde{y}_1^k)$, $p(\tilde{y}_1^k)$ and MSE (6) can be determined [3,4] using model (4) and known distributions of the signal and noises.

Unlike Bayesian measures commonly used in optimization of digital units of the receivers, MSE (6) depends not only on the algorithm $\hat{x}_k = \hat{x}_k(\tilde{y}_1^k)$ of estimates computing but also on the parameters $B_1^k, M_1^k$ of adaptive modulator M1, as well as on its possible over-modulation (saturation). Explicit dependence of MSE (6) on the parameters of transmitting and receiving parts of ACS allows us to give following, initial formulation of full optimization task:

Initial formulation: one should find the estimation algorithm $\hat{x}_k = \hat{x}_k(\tilde{y}_1^k)$ and controls $M_k = M_k(\tilde{y}_1^{k-1})$, $B_k = B_k(\tilde{y}_1^{k-1})$ which minimize MSE (6) for each $k = 1, ..., n$.

Direct solution of this task is impossible due to invincible mathematical difficulties caused by saturation form of modulator characteristic (2). However, this task becomes solvable, if additional measure of the transmission quality – permissible probability $\mu$ of over-modulation is introduced.

#### A. Statistical fitting condition [3-5]

Definition: adaptive modulator is *statistically fitted* to the input signal if, for each $k = 1, ..., n$, its parameters $B_k, M_k$ satisfy the inequality:

$$\Pr_k^{o.m.} = \Pr(M_k |x_k - B_k| > 1 \mid \tilde{y}_1^{k-1}, B_1^{k-1}, M_1^{k-1}) < \mu \quad (7)$$

called the *statistical fitting condition*.

This condition can be written in the equivalent and more convenient for practical calculations form:

$$\Pr_k^{lin} = \int_{B_k - \frac{1}{M_k}}^{B_k + \frac{1}{M_k}} p(x_k | \tilde{y}_1^{k-1}) \, dx \geq 1 - \mu \quad (8)$$

Values $\Pr_k^{o.m.}$ and $\Pr_k^{lin}$ in (7),(8) are the probabilities of over-modulation and of linear mode of the modulator M1 work in $k$-th cycle, respectively. Value $\mu$ determines the permissible probability of over-modulation in each cycle of the sample transmission. Depending on the requirements to the system (specified by designers), in most of practical cases values $\mu$ lay in the interval $10^{-4} \leq \mu < 10^{-12}$.

Definition: inequalities (7),(8) determine the set $\Omega_k$ of *permissible values* of the parameters $\{B_k, M_k\}$ which do not violate these inequalities.

In Appendix A, we show that over-modulation in "pre-threshold" (see Sect. IV) cycles of transmission not only corrupts the observation but also violates fitting condition (7) that radically diminishes a probability of restoration of the linear mode of transmission. This results in appearance of abnormal errors in final estimates $\hat{x}_n$. A probability of first appearance of over-modulation in $k$-th cycle has the value: $\mu + O[(k-1)\mu^2]$ and determines the mean percent of erroneous estimates in sequences of estimates at the AFCS output. Assuming that undistorted estimate $\hat{x}_n$ delivers amount of information $I(X, \hat{X}_n)$ bits, one may consider $\mu$ as the mean percent of distorted bits or the probability of appearance of erroneous bit in information flow at the AFCS output.

Intermediate conclusion: probability $\mu$ is the characteristic of AFCS fidelity similar to the bit-error-rate (BER) in digital transmission systems.

### B. Formulation of "solvable" full optimization task

The statistically fitted modulator works practically always (with probability $1-\mu$, for each $k$) as the linear unit. In this case, non-linear model of transmitter (2) can be replaced by the linear one:

$$s_{t,k} = A_0 M_k (x_k - B_k) \cos(2\pi f_0 t + \varphi_0) . \quad (9)$$

and model (4) of the signal at the DSPU input can be written in the form:

$$\tilde{y}_k = A M_k (x_k - B_k) + \xi_k . \quad (10)$$

The results obtained using formulas (2) and (9) will differ by the values of $O(\mu)$ order.

Transition to the linear model (10) permits to find explicit form of MSE (6) and reduce initially non-linear optimization task to the linear "solvable" one which can be formulated as follows:

Final formulation of optimization task for AFCS:

For each $k = 1,..,n$, one should find estimates $\hat{x}_k = \hat{x}_k(\tilde{y}_1^k)$ and controls $B_k = B_k(\tilde{y}_1^{k-1})$, $M_k = M_k(\tilde{y}_1^{k-1})$ which minimize MSE (6), constructed using linear model (10), under fulfilled statistical fitted condition (7).

This task can be solved using known methods of Bayesian estimation theory in two steps [3-5]. Namely:

Step (A): assuming the adjusted parameters of modulator M1 take only permissible values: $\{B_k, M_k\} \in \Omega_k$, one should find the estimation algorithm $\hat{x}_k = \hat{x}_k(\tilde{y}_1^k, B_1^k, M_1^k)$ minimizing MSE (6) for each cycle of the sample transmission.

Step (B): found in Step (A) optimal estimate $\hat{x}_k(\tilde{y}_1^k, B_1^k, M_1^k)$ should be substituted into MSE (6), and the result of substitution should be minimized over $\{B_k, M_k\} \in \Omega_k$.

In [3,4], general solution of full optimization task based on the posterior PDF is given. Below, simpler solution is presented. The analysis is carried out under assumption that fitting condition (8) is fulfilled for each cycle of the sample transmission, and sources of signals and noises are Gaussian.

### C. Solution of optimization task

In Gaussian case, optimal Bayesian estimates minimizing MSE (6) are linear combinations of observations [16],[17]. This allows us to search them using equation (5) that reduces the task to the search of the values $L_k$ minimizing MSE (6). Substituting (10) into (5) and subtracting value $x$ from both sides of equation, then taking into account the equality $E(\tilde{y}_k | \tilde{y}_1^{k-1}) = AM_k(\hat{x}_{k-1} - B_k)$, one may obtain the relationship:

$$\hat{x}_k - x = (1 - AM_k L_k)(\hat{x}_{k-1} - x) + L_k(AM_k v_k + \xi_k) . \quad (11)$$

Averaging the squared equation (11) gives the recurrent equation for MSE:

$$P_k = (1 - AM_k L_k)^2 P_{k-1} + L_k^2 (A^2 M_k^2 \sigma_v^2 + \sigma_\xi^2) \quad (12)$$

with the initial condition $P_0 = \sigma_0^2$.

Minimizing (12) over the gains $L_k$ under given $P_{k-1}$ and permissible values of parameters $B_1^{k-1}, M_1^{k-1} \in \Omega_1 \times ... \times \Omega_{k-1}$, (Step (A) of AFCS optimization) gives following expression for optimal gains:

$$L_k = \frac{AM_k P_{k-1}}{\sigma_\xi^2 + A^2 M_k^2 (\sigma_v^2 + P_{k-1})} . \quad (13)$$

Substitution of (13) into (12) gives the equation determining low boundary of MSE values on the set of linear estimation algorithms:

$$P_k = \frac{(\sigma_\xi^2 + A^2 M_k^2 \sigma_v^2) P_{k-1}}{\sigma_\xi^2 + A^2 M_k^2 (\sigma_v^2 + P_{k-1})} = P_{k-1} - \frac{A^2 M_k^2 P_{k-1}^2}{\sigma_\xi^2 + A^2 M_k^2 (\sigma_v^2 + P_{k-1})} \quad (14)$$

According to (14) MSE of transmission decreases for the greater values of modulation depth $M_k$ and does not depend on the controls $B_k$. The latter simplifies Step (B) of optimization, which can be reduced to the search of maximal value $M_k$ in the set $\Omega_k$ determined by condition (8). In Gaussian case, this condition takes the form:

$$\int_{B_k - \frac{1}{M_k}}^{B_k + \frac{1}{M_k}} p(x_k | \tilde{y}_1^{k-1}) dx_k = \frac{1}{\sqrt{2\pi(\sigma_v^2 + P_{k-1})}} \int_{B_k - \frac{1}{M_k}}^{B_k + \frac{1}{M_k}} e^{-\frac{[x - E(x_k | \tilde{y}_1^{k-1})]^2}{2(\sigma_v^2 + P_{k-1})}} dx \geq 1 - \mu \quad (15)$$

where $E(x_k | \tilde{y}_1^{k-1}) = E(x + v_k | \tilde{y}_1^{k-1}) = \hat{x}_{k-1}(\tilde{y}_1^{k-1})$ is the predicted value of the sample which coincides with optimal estimate of the sample formed in previous cycle of transmission.

One may notice that maximal satisfying (15) values $M_k$ depend on $B_k$ and lay on the boundary of conditional set $\Omega_k\{|B_k\}$. Maximal in $\Omega_k\{|B_k\}$ value $M_k$ refers to the point

$$B_k = E(x_k | \tilde{y}_1^{k-1}) = \hat{x}_{k-1}(\tilde{y}_1^{k-1}) = \hat{x}_{k-1} . \quad (16)$$

Substituting (16) into (15) gives the equation:

$$\frac{1}{\sqrt{2\pi(\sigma_v^2 + P_{k-1})}} \int_{-\frac{1}{M_k}}^{\frac{1}{M_k}} e^{-\frac{x^2}{2(\sigma_v^2 + P_{k-1})}} dx = 1 - \mu \quad (17)$$

which can be rewritten in the form:

$$\Phi(\alpha) = \frac{1}{\sqrt{2\pi}} \int_0^\alpha e^{-\frac{x^2}{2}} dx = \frac{1 - \mu}{2} \quad (18)$$

where $\Phi(\alpha)$ is known Gaussian error function. Parameter $\alpha = 1/M_k(\sigma_v^2 + P_{k-1})$ can be called the *over-modulation* or *saturation factor*. The latter relationship and (18) determine optimal, for each cycle, values of modulation depth:

$$M_k = \frac{1}{\alpha \sqrt{\sigma_v^2 + P_{k-1}}} \quad (19)$$

which do not depend on observations.

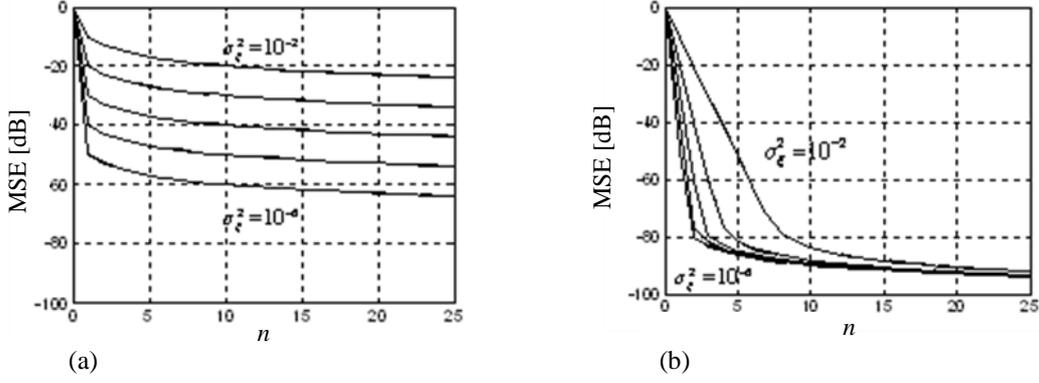

Fig. 4. Dependencies of MSE of the sample estimates on the number of cycles under different values of the forward channel noise power $\sigma_\xi^2$: a) in optimal analogue CS without feedback ($B_k = x_0$, $M_k = M_1$); b) in optimal AFCS.

## VI. PARTICULARITIES OF OPTIMAL AFCS WORK

According to known results [16],[17], in the Gaussian case, estimates $\hat{x}_{k-1} = E(x_k | \tilde{y}_1^{k-1})$ orthogonalize the residuals $e_k = x_k - \hat{x}_{k-1}$. Therefore, the signal $x_k - \hat{x}_{k-1}$ at the input of modulator M1 becomes the zero mean white Gaussian noise with the power $\sigma_\nu^2 + P_k$ monotonically diminishing for greater number of cycles. In turn, increase of $M_k$ according to (19) makes the emitted by M1 signal (2) the stationary Gaussian pulse-amplitude modulated signal of the white-noise-type (sequence of independent Gaussian PAM pulses). Each pulse is transmitted during the time $\Delta t_0 = 1/2F_0$ in the frequency band $[f_0 - F_0, f_0 + F_0]$. Mean power of the pulses at the channel Ch1 output (demodulator DM1 input) attains maximal, under given $\mu$, value:

$$W_k^{sign} = E(s_{t,k}^2) = A^2 M_k^2 (\sigma_\nu^2 + P_{k-1}), \quad (A = A_0 \gamma / r) \quad (20)$$

(coefficient $1/2$ is included into $\gamma$). In this case, observations $\tilde{y}_1^k = (\tilde{y}_1, ..., \tilde{y}_k)$ at the DM1 output become the stationary white Gaussian noise with the zero mean value:

$$E(\tilde{y}_k | \tilde{y}_1^{k-1}) = AM_k E(x_k - B_k | \tilde{y}_1^{k-1}) = AM_k (\hat{x}_{k-1} - B_k) = 0 \quad (21)$$

and variance:

$$E(\tilde{y}_k^2 | \tilde{y}_1^{k-1}) = \sigma_\xi^2 + A^2 M_k^2 (\sigma_\nu^2 + P_{k-1}) = \sigma_\xi^2 (1+Q^2) \quad (22)$$

which can be easily found by averaging of squared formula (10). Value $Q^2$ in (22) denotes maximal available signal-to-noise ratio (SNR) at the DSPU input (demodulator DM1 output):

$$Q^2 = \frac{W_k^{sign}}{W^{noise}} = \frac{A^2 M_k^2 (\sigma_\nu^2 + P_{k-1})}{\sigma_\xi^2} = \left(\frac{A_0 \gamma}{r\alpha}\right)^2 \frac{1}{N_\xi F_0} \quad (23)$$

Summary. The obtained results determine the final form of optimal transmission/receiving algorithm. Its particularity is the mixed-signal structure determining joint optimal work of the analogue and digital units of AFCS:

Optimal algorithm of AFCS functioning:

*1. Optimal algorithm of estimates computing (unit DSPU):*

Application of (21) to recursion (5) gives the equation for optimal estimates computing:

$$\hat{x}_k = \hat{x}_{k-1} + L_k \tilde{y}_k . \quad (24)$$

where observations $\tilde{y}_k$ are formed according to (10), and the gains $L_k$ have the values:

$$L_k = \frac{AM_k P_{k-1}}{\sigma_\xi^2 + A^2 M_k^2 (\sigma_\nu^2 + P_{k-1})} = (AM_k)^{-1}\left(1 - \frac{P_k}{P_{k-1}}\right) \quad (25)$$

that can be easily checked using formulas (13),(14). In turn, substitution of (19) into (14) gives the relationship:

$$P_k = (1+Q^2)^{-1}\left[1 + Q^2 \frac{\sigma_\nu^2}{(\sigma_\nu^2 + P_{k-1})}\right] P_{k-1} . \quad (26)$$

Initial conditions for recursions (24),(26): $\hat{x}_0 = x_0$; $P_0 = \sigma_0^2$.

*2. Optimal controls for the analogue modulator adjusting:*

Parameters of modulator described by model (2) should be set, for each $k = 1,...,n$, to the values:

$$B_k = \hat{x}_{k-1}(\tilde{y}_1^{k-1}) = \hat{x}_{k-1}; \quad M_k = \frac{1}{\alpha \sqrt{\sigma_\nu^2 + P_{k-1}}} \quad (27)$$

One should notice that inversion of the sign of analogue gains $M_k$ does not influence the minimal MSE (MMSE), i.e. the quality of transmission that can be used in practical AFCS design.

Intermediate conclusion: equation (26) determines absolute low boundary of possible values of MSE of formed by AFCS estimates under given probability $\mu$. Relationships (24)-(27) give all necessary information for designing the AFCS working at this boundary.

Analysis of dependence of (26) on the number of cycles shows that at the initial interval $1 < k \le n^*$, where $P_{k-1}/\sigma_\nu^2 \gg 1 + Q^2$, MMSE diminishes exponentially:

$$P_k = \sigma_0^2 (1+Q^2)^{-k} . \quad (28)$$

After "threshold" number of cycles $n^*$ determined by the equation $P_{n^*} = \sigma_\nu^2$, this dependence takes hyperbolical form:

$$P_k = \frac{\sigma_\nu^2}{k - n^* + 1} \quad (29)$$

that follows form (26) under $P_{k-1}/\sigma_\nu^2 \ll 1 + Q^2$.

The threshold number of cycles $n^*$ can be evaluated by substitution of (28) into equation $P_{n^*} = \sigma_\nu^2$ that gives the assessment:

$$n^* = \frac{1}{\log_2(1+Q^2)} \log_2\left(\frac{\sigma_0^2}{\sigma_\nu^2}\right). \quad (30)$$

Improvement of transmission quality in optimal AFCS in comparison with optimal CS without feedback is illustrated in Fig. 4. Presented in Fig. 4a plots were computed using

(14) under condition: $M_k = M_1 = 1/\sqrt{\sigma_v^2 + \sigma_0^2}$. In this case, MMSE of estimates, as a function of $k$, diminishes hyperbolically independently from the number of cycles:

$$P_n = \frac{\sigma_0^2}{1 + nQ_{CS}^2} \quad (31)$$

where SNR $Q_{CS}$ at the channel output should be computed for the bandwidth $2F$ (usual PAM does not extend the spectrum of input signals), that is

$$Q_{CS} = \frac{W^{sign}}{N_\xi F} = \frac{A^2 M_1^2 \sigma_0^2}{N_\xi F} =$$
$$= \left(\frac{A_0 \gamma}{r\alpha}\right)^2 \frac{1}{N_\xi F} \frac{\sigma_0^2}{(\sigma_v^2 + \sigma_0^2)}\bigg|_{\sigma_0^2 \gg \sigma_v^2} = \left(\frac{A_0 \gamma}{r\alpha}\right)^2 \frac{1}{N_\xi F} + O\left(\frac{\sigma_v^2}{\sigma_0^2}\right). \quad (32)$$

Plots in Fig. 4b for MMSE of optimal AFCS were computed using (26) under the same parameters ($A$=1.25; $\alpha$=4; $\sigma_0$=1.25; $\sigma_v^2 = 10^{-8}\sigma_0^2$). Points in Fig. 4a correspond to the threshold numbers of cycles. Simulation experiments with full models of the systems gave dependences practically identical to theoretical plots.

Exponentially fast diminution of MMSE of transmission at the interval $1 < k \leq n^*$ is the effect conditioned by joint optimal adaptive adjusting of TU and processing of observations. This effect was studied in earlier works on the analogue communication by Goblick [9], Kailath [8], Omura [10,11], Schalkwijk and Bluestein [12], Gallager [13] and other authors [2] under limitation on mean power of the emitted signals. In later researches, optimization task of analogue FCS with Markov sources and noiseless feedback channel was considered by Baccarelly and Cusani [14], where relationships similar to (28), (30) were obtained as the particular case. Further development and application of these results, as it was noted in Introduction, was hampered by the assumed linearity of the transmitter. Apart of difficulties with practical implementation of theoretical results, this does not permit to analyze the effects appearing under $k > n^*$, when MMSE $P_k$ of estimates $\hat{x}_k$ becomes less than the threshold value $P_{n^*} = \sigma_v^2$.

<u>Remark:</u> for noiseless feedback channels ($\sigma_v^2 = 0$), MMSE of transmission diminishes exponentially for arbitrary number of cycles that was shown in the quoted works. This is valid also for considered optimal AFCS. Under $\sigma_v^2 = 0$, formulas (14), (19) take the form:

$$P_k = \frac{\sigma_\xi^2 P_{k-1}}{\sigma_\xi^2 + A^2 M_k^2 P_{k-1}} \; ; \quad M_k = \frac{1}{\alpha \sqrt{P_{k-1}}} \quad (33)$$

Substituting $M_k$ into $P_k$ gives necessary result:

$$P_k = \frac{\sigma_\xi^2 P_{k-1}}{\sigma_\xi^2 + \frac{A^2}{\alpha^2}} = P_{k-1}(1 + Q^2)^{-1} = \sigma_0^2 (1 + Q^2)^{-k} \quad (34)$$

This result can be obtained directly from (26).

IV. INFORMATION CHARACTERISTICS OF OPTIMAL AFCS

Further analysis will be held under assumptions: AFCS is statistically fitted to the input signal and operates according to optimal algorithm (24)-(27). Each sample is transmitted in $n$ cycles in the same way independently from the previous samples, and duration of a single cycle of transmission $\Delta t_0 = 1/2F_0$ (channel bandwidth $2F_0$) is fixed.

A. *Bit-rate of transmission in the channel M1-Ch1-DM1*

Substitution of (27) into (10) to write the signals at the DSPU unit in the form

$$\tilde{y}_k = AM_k(x_k - \hat{x}_{k-1}) + \xi_k \quad (35)$$

valid for each $1 \leq k \leq n$ independently of $n$. Using (35) and (21), (22), one may easily obtain following relationships for conditional PDFs:

$$p(\tilde{y}_k, x_k | \tilde{Y}_1^{k-1}) = \frac{1}{\sqrt{2\pi\sigma_\xi^2}} \exp\left(-\frac{[y_k - AM_k(x_k - \hat{x}_{k-1})]^2}{2\sigma_\xi^2}\right) \quad (36)$$

$$p(\tilde{y}_k | \tilde{Y}_1^{k-1}) = \frac{1}{\sqrt{2\pi\sigma_\xi^2(1+Q^2)}} \exp\left(-\frac{\tilde{y}_k^2}{2\sigma_\xi^2(1+Q^2)}\right) \quad (37)$$

and corresponding entropies:

$$H(\tilde{Y}_k | X_k, \tilde{Y}_1^{k-1}) = \frac{1}{2}\log_2(2\pi e \sigma_\xi^2) \quad (38)$$

$$H(\tilde{Y}_k | \tilde{Y}_1^{k-1}) = \frac{1}{2}\log_2[2\pi e \sigma_\xi^2(1+Q^2)] \quad (39)$$

Then, amount of information $I(\tilde{Y}_k, X_k | \tilde{Y}_1^{k-1})$ in observation $\tilde{y}_k$ about the signal $x_k$ at the input of the modulator M1 has the value:

$$I(\tilde{Y}_k, X_k | \tilde{Y}_1^{k-1}) = H(\tilde{Y}_k | \tilde{Y}_1^{k-1}) - H(\tilde{Y}_k | X_k, \tilde{Y}_1^{k-1}) =$$
$$= \frac{1}{2}\log_2(1+Q^2) = \frac{1}{2}\log_2\left(1 + \frac{W^{sign}}{N_\xi F_0}\right) \text{ [bit/cycle]}. \quad (40)$$

Taking into account duration of the single cycle $\Delta t_0 = 1/2F_0$ and (40), one may find mean bit-rate of transmission through the channel M1-Ch1-DM1:

$$R_k^{Ch1} = \frac{I(\tilde{Y}_k, X_k | \tilde{Y}_1^{k-1})}{\Delta t_0} = F_0 \log_2\left(1 + \frac{W^{sign}}{N_\xi F_0}\right) \text{ [bit/s]} \quad (41)$$

Formula (41) coincides with Shannon's formula for the capacity of the channels with AGWN, that is

$$R_k^{Ch1} = R_{\max}^{Ch1} = C = F_0 \log_2\left(1 + \frac{W^{sign}}{N_\xi F_0}\right) \quad (42)$$

and, that is important, is constant independently from the number of cycles of the sample transmission.

Single particularity differing (42) from the basic expression for the Gaussian channel capacity is the method of the mean power of the received signal computing. In AFCS, it has the form: $W^{sign} = (A/\alpha)^2$ where saturation factor $\alpha$ is determined by (18) and depends on the permissible (assumed) probability of over-modulation (BER) $\mu$.

<u>Intermediate conclusion</u>: The obtained result shows that optimal adjusting the modulator M1 according to (27) increases a bit-rate up to the capacity of the channel M1-Ch1-DM1 independently from a quality of the feedback channel. This effect is a result of maintaining the power of emitted signal at the maximal, under given BER $\mu$, and constant for each $k$ value $A_0^2/2\alpha^2$. Achievement of the greater bit-rate is impossible. This result shows that optimal AFCS completely employs the resources of the forward channel.

## B. Bit-rate at the AFCS output (rate-distortion function)

Let each sample be transmitted in $n$ cycles in the same way independently from the previous ones, channel bandwidth $2F_0$ is fixed and $T = n\Delta t_0 = n/2F_0$. Amount of information $I(X, \hat{X})$ in the final estimate $\hat{x}_n$ about the sample $x$, in the Gaussian case, can be easily computed and is determined by the formula(e.g. [8], [13], [16]):

$$I(X, \hat{X}_n) = \frac{1}{2}\log_2 \frac{\sigma_0^2}{P_n} \quad (43)$$

and the bit-rate at the AFCS output takes the values:

$$R_n^{AFCS} = \frac{I(X, \hat{X}_n)}{T} = 2FI(X, \hat{X}_n) = \frac{F_0}{n}\log_2 \frac{\sigma_0^2}{P_n}. \quad (44)$$

On the other hand, formulas (43),(44) present the Shannon's distortion and rate distortion functions for Gaussian source [7]-[12] and determines minimal information per sample and mean bit-rate necessary for AFCS transmit the input samples $x$ with MSE not smaller than $P_n$.

Let us notice that formulas (43), (44) are valid for arbitrary, not necessary optimal (but statistically fitted) Gaussian AFCS. In the optimal systems built according to (24)-(27), MSE $P_n$ of estimates $\hat{x}_n$ attains, for each $n$, minimal value (26). In this case, distortion rate (44) determines *maximal* bit-rate achievable at the AFCS output on the set of estimation algorithms which can be used for the output estimates $\hat{x}_n$ computing. From our point of view, according to sense of this value, rate distortion function (44) with MMSE $P_n$ can be considered the as the *capacity of the system* as a unit.

Formalizing this conclusion, one may consider AFCS as a specific "macro" channel described by the relationship:

$$\hat{x}_n = x + \eta_n \quad (45)$$

where $x$ is the input signal with known PDF $p(x)$ that is necessary condition for the MSE and fitting condition definition, $\hat{x}_n$ is output signal. Unlike "usual" Gaussian channels, additive Gaussian noise $\eta_n$ (describing estimation errors) has the variance $E(\eta_n^2) = E[(x - \hat{x}_k)^2]$ which depends on the algorithm of estimation. Therefore, maximal bit-rate (capacity of AFCS as communication unit) should be searched on the set of possible algorithms of estimation under given PDF $p(x)$, and can be determined as follows:

$$C^{syst} = F_0 \max_{\hat{x}_k = \hat{x}_k(\tilde{y}_1^k)} \frac{I(X, \hat{X}_n)}{n} = \frac{F_0}{n} \max_{\hat{x}_k = \hat{x}_k(\tilde{y}_1^k)} \log_2 \frac{\sigma_0^2}{P_n}. \quad (46)$$

The extreme in (46) should be searched under the same additional constraints as in the full optimization task.

According to (28),(29), MMSE of the output estimates $\hat{x}_n$ and, as a consequence, bit-rate (44) depend on the number of cycles $n$ in different way. Substitution of (28),(29) into (44) gives the following results:

- termination of the sample transmission in "pre-threshold" interval $1 \leq n \leq n^*$ results in MMSE $P_n$ is determined by formula (28), and formula (44) takes the form:

$$R_n^{AFCS} = F_0 \log_2(1 + Q^2) = F_0 \log_2\left(1 + \frac{W^{sign}}{N_\xi F_0}\right) \quad (47)$$

i.e. is equal to the capacity $C$ of the forward channel.

- termination of the transmission in the "post-threshold" interval $n > n^*$ results in formula (44) taking the form:

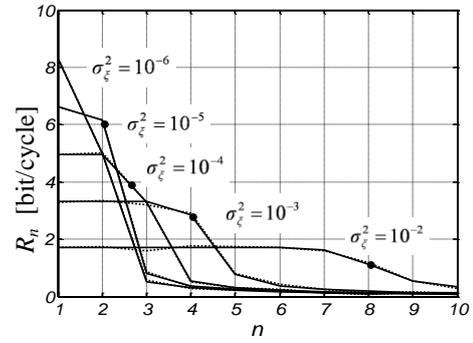

Fig. 3. Changes of the bit-rate at AFCS output under different power $\sigma_\xi^2$ of the forward channel noise ($\alpha = 4, \sigma_0 = 1.25, \sigma_v^2 = 10^{-8}\sigma_0^2, A = 1.25$), continuous lines refer to the theoretical dependencies, dotted lines – to the empirical ones, points denote the assessments of the threshold number of cycles.

$$R_n^{AFCS} = \frac{F_0}{n}\left[\log_2\left(\frac{\sigma_0^2}{\sigma_v^2}\right) + \log_2(n - n^* + 1)\right] \quad (48)$$

and monotonically diminishes from the value $C$ to zero.

In Fig. 3, dependencies $R_n^{AFCS} = R(n)$ of maximal bit-rate at the AFCS output under different values of the power of channel noise $\sigma_\xi^2$ are shown (continuous lines). Also, empirical dependencies $\hat{R}_n^{AFCS} = \hat{R}(n)$ obtained in simulation experiment are presented (dotted lines). The system was modeled using relationships (4), (24)-(27) and digitally generated Gaussian samples and noises. The bit-rate at its output was computed using (43) with $P_n$ replaced by the empirical MMSE $\hat{P}_n$:

$$\hat{P}_n = \frac{1}{M}\sum_{m=1}^{M}\left[x^{(m)} - \hat{x}_n^{(m)}\right]^2, \quad (49)$$

where $x^{(m)}$ are elements of Gaussian sequence of the samples $\{x^{(1)},...,x^{(M)}\}$, and $\hat{x}_n^{(m)}$ - their estimates, $M = 5000$ samples.

Constant and equal to the channel capacity value of bit-rate $R_n^{AFCS}$ in the interval $1 < n \leq n^*$ is a result of practically full suppression of the forward channel noise $\xi_k$ due to fast (exponential) increase of the modulation depth $M_k$ in pre-threshold interval $1 < n \leq n^*$. After $n$ cycles of transmission, this creates the effect equivalent to $M_1 \times ... \times M_n$ times increase of SNR at the channel M1-Ch1-DM1 output, although mean power of the emitted signal remains constant for each cycle of transmission. At the $n^*$-th cycle, power of informative component $x - \hat{x}_n$ in the signal $e_k$ at the modulator M1 input attains the level of the noise $v_k$ power:

$$E[(x - \hat{x}_{n^*-1})^2] = P_{n^*} = \sigma_v^2. \quad (50)$$

For this reason, beginning with $n > n^*$, noise in the forward channel practically does not influence the bit-rate $R_n^{AFCS}$ which depends now only on the SNR $\sigma_0^2/\sigma_v^2$. Slow diminution of MMSE $P_k$ in the post-threshold interval $n^* < k \leq n$ is a result of continued estimation of small and monotonically diminishing difference signal $x - \hat{x}_n$ in relatively "powerful" noise $v_k$. For this reason, although the bit-rate in the forward channel remains equal to its capacity, percent of useful information in delivered to DSPU observations $\tilde{y}_n$ monotonically decreases. This results in decrease of information flow at the AFCS output and diminishes the capacity of the system.

Summary: The obtained results show:

1. Bit-rate (44) in optimal AFCS attains maximal value and can be considered as the capacity of the system as the unit.

2. Optimal AFCS fully employ the information recourses of their transmitting and receiving units and transmit the signals through the forward channel Ch1 with the bit rate equal to its capacity (42) which does not depend on the feedback channel characteristics and number of transmission cycles.

3. Capacity (48) of optimal AFCS is constant and equal to capacity of the forward channel, if the number of transmission cycles $n$ is does not exceed the threshold number $n^*$. For $n > n^*$ capacity of AFCS does not depend on the forward channel characteristics and monotonically diminishes.

4. In the pre-threshold interval $1 \leq n \leq n^*$, amount of information delivered to addressee grows linearly at bit-rate (47), and in post-threshold interval $n > n^*$ - logarithmically at bit-rate (48).

5. Increase of the number of transmission cycles per sample, under fixed bandwidth of the feed-forward channels results in necessary narrowing of the system bandpass ($F = F_0/n$). Reverting this statement, one may say that transmission of signals in the baseband [-F, F] in $n$ cycles requires the channels bandwidth should be extended by $n$ times ($F_0 = nF$).

6. Feedback channel noise does not influence the capacity of optimal AFCS in the interval $1 \leq n \leq n^*$ but determines the threshold number of cycles (30). According to (30), the greater feedback noise power, the shorter threshold interval. For

$$\sigma_v^2 = \sigma_0^2 (1 + Q^2)^{-1} \qquad (51)$$

threshold effect disappears: $n^* = 1$, and AFCS work as almost non-adaptive system with the output bit-rate

$$R_n^{AFCS} = \frac{F_0}{n} \left[ \log_2(1+Q^2) + \log_2 n \right] \qquad (52)$$

It is interesting that optimal AFCS continues to work even if the feedback noise power be greater than the power of input signals. For $\sigma_v^2 >> \sigma_0^2$ formula (26) takes the form:

$$P_n = \frac{\sigma_v^2 P_{n-1}}{\sigma_v^2 + P_{n-1}} = \sigma_0^2 \left(1 + n\frac{\sigma_0^2}{\sigma_v^2}\right)^{-1} \qquad (53)$$

However, corresponding bit-rate is very small and diminishes beginning with the first cycle of transmission.

$$R_n^{AFCS} = \frac{F_0}{n} \log_2\left(1 + n\frac{\sigma_0^2}{\sigma_v^2}\right) \qquad (54)$$

V. EFFICIENCY OF AFCS WORK AND APPLICATIONS

In Sect. IV, it was shown that bit-rate at the optimal AFCS output quickly diminishes after threshold number of cycles $n > n^*$, although MMSE of the output estimates continue to diminish, and amount of information in estimates $\hat{x}_n$ grows. This means that full optimization of AFCS does not solve all the questions, and next group of questions appears concerning - how these systems should be applied to ensure the most efficient utilization of their resources. First of all, what number of cycles ensures most efficient work of the system?

*A. Threshold systems*

One can easily check that, in the post-threshold interval $n > n^*$, each additional bit of information in estimates $\hat{x}_n$ can be obtained in not less than $3(n - n^* + 1)$ additional cycles of transmission [4], while in the pre-threshold interval each cycle adds $1/2 \cdot \log_2(1+Q^2)$ significant bits to the estimate.

For that reason, each additional bit obtained in post-threshold interval is tied with much greater narrowing of the AFCS bandpass $F = F_0/n$ (or corresponding extension of the channels bandwidth), than it is in the pre-threshold interval.

In turn, termination of transmission inside of the pre-threshold interval $1 \leq n < n^*$) extends the band-pass of AECS $F = F_0/n$ but increases MMSE, in comparison with the potentially achievable during $n^*$ cycles of transmission up to the value:

$$P_n = P_{n^*}(1+Q^2)^{n^*-n} = \sigma_v^2(1+Q^2)^{n^*-n} = \overline{\sigma}_v^2. \qquad (55)$$

The latter means not complete utilization of resources of given AFCS. Namely, if the accuracy $P_n = \overline{\sigma}_v^2$ is sufficient for external addressee, one may significantly weaken the requirements to the level of analogue noise $\overline{v}_k$ at the input of modulator M1. This means not only a possibility to use, in TU, the simpler elements with greater internal noise but also to weaken the requirements to the quality of feedback channel M2-Ch2-DM2.

Intermediate conclusion: most efficient "compromise solution" is transmission of the sample in $n^*$ cycles. This is maximal number of cycles permitting to deliver information to addressee with bit-rate equal to the capacity of the forward channel (of AFCS). Shorter transmission causes not full utilization of the system's resources, longer – sharp diminution of the bit-rate (capacity of AFCS) and of mean number of bits delivered in each next cycle of transmission.

Not less important is connection between the number of cycles and power efficiency of transmission. Energy of the signal received by BS during $n$ cycles of the sample transmission is determined by the formula:

$$E_n^{sign} = W^{sign} n \Delta t_0 = n \frac{W^{sign}}{2F_0} \qquad (56)$$

($W^{sign}$ is the power of received signal). Then, energy per bit at the AFCS can be assessed by the relationship:

$$E^{bit} = \frac{E_n^{sign}}{I(X,\hat{X}_n)} = \frac{W^{sign} n}{F_0 \log_2\left(\frac{\sigma_0^2}{P_n}\right)} =$$

$$= \begin{cases} \dfrac{W^{sign}}{F_0 \log_2(1+Q^2)} = \dfrac{N_\xi Q^2}{\log_2(1+Q^2)} & \text{for } 1 \leq n \leq n^* \\ \dfrac{W^{sign} n}{F_0 \left[\log_2\left(\dfrac{\sigma_0^2}{\sigma_v^2}\right) + \log_2(n - n^* + 1)\right]} & \text{for } n > n^* \end{cases} \qquad (57)$$

where (23) is taken into account (i.e. $Q^2 = W^{sign}/N_\xi F_0$).

Formula (57) shows that transmission of the sample in $n > n^*$ cycles causes fast growth of the energy per bit. For $1 \leq n < n^*$, this value is constant, but as it was shown above, resources of the AFCS are utilized not completely. These effects also confirm the conclusion that $n^*$ is most efficient, "optimal" number of the cycles of the sample transmission.

Summary: Most efficient way of optimal AFCS application is the "*threshold mode*", when the number of cycles of the sample transmission is equal to the threshold number $n = n^*$ (30). In the interval $[1; n^*]$, AFCS delivers information to addressee with a bit-rate equal to the capacity of forward channel, and the system works completely utilizing resources of its transmitting and receiving parts (TU and BS).

Definition: optimal AFCS used in the threshold mode are called further "threshold systems". The characteristics of this group of systems are described by only the part of obtained above relationships that are valid in the interval $1 \leq n \leq n^*$.

Let us notice that variance $\sigma_v^2$ includes, as the component, variance of estimation errors at the feedback channel output. This allows us to formulate

Intermediate conclusion: MMSE of estimates at the output of threshold AFCS is always not smaller than MSE of estimates delivered to TU (to the input of modulator M1) through the feedback channel.

*B. Efficiency of the threshold AFCS*

According to definition, $n^* = F_0 / F$ (30) takes the form of known in information theory relationship:

$$R_{n^*} = F \log_2\left(\frac{\sigma_0^2}{\sigma_v^2}\right) = F_0 \log_2\left(1 + \frac{W^{sign}}{N_\xi F_0}\right) = C \quad (58)$$

which describes the most efficient bandwidth-SNR tradeoff in *ideal* communication systems. In turn, value $n^* = F_0 / F$ establishes optimal, under given conditions, relation between the baseband of the input signal and bandwidth of the forward channel Ch1 (also of the channel Ch2).

Intermediate conclusion: formula (58) follows directly from algorithm (24)-(27). This permits to claim that AFCS built using (24)-(27) and applied in the threshold mode work most efficiently, as ideal communication systems.

Taking into account that, in threshold AFCS, MMSE of output estimates $P_{n^*}$ attains the value $P_{n^*} = \sigma_v^2 = \sigma_{out}^2$, formula (58) can be rewritten in the form of relationship:

$$\sigma_{out}^2 = \sigma_0 \left(1 + Q^2\right)^{-\frac{F_0}{2F}} = \sigma_0\left(1 + \frac{W^{sign}}{N_\xi F_0}\right)^{-\frac{F_0}{2F}} = \sigma_0^2 e^{-\frac{C}{F_0}} \quad (59)$$

which determines minimal absolute and relative errors of the sample transmission by threshold AFCS and coincides with corresponding formulas in [8]-[13].

One may add that fulfilment of the condition $P_{n^*} = \sigma_{out}^2 = \sigma_v^2$ means that, in the threshold mode, $SNR_{out}$ at the AFCS output is equal to $SNR_{inp}$ at the input of modulator M1:

$$SNR_{inp} = \frac{\sigma_0^2}{\sigma_v^2} = SNR_{out} = \frac{\sigma_0^2}{\sigma_{out}^2} \quad (60)$$

Defining "normalized" SNR at the AFCS output as $E^{bit} / N_x$, and channel bandwidth efficiency as $C/F_0$ one may rewrite (57) in the form more convenient for analysis of the system efficiency [1]:

$$\frac{E^{bit}}{N_\xi} = \frac{F_0}{C}\left(2^{\frac{C}{F_0}} - 1\right). \quad (61)$$

This dependence is presented in Fig. 5 (also in Fig. 1) as the line $R = C$ (Shannon's boundary, in generalization of the term "Shannon limit" $E^{bit}/N_x = -1,6$ dB under $C/F_0 \to \infty$). The line divides the efficiency plane "bandwidth/SNR" (in dB) into "unachievable" region ($R > C$), and the region of real systems ($R < C$). Main task of contemporary theory and practical design is to approach efficiency of CS – computed or measured points ($E^{bit}/N_x$, $R/F_0$) - to the Shannon's boundary. As illustration, in Fig. 1, power-bandwidth tradeoff in CS with coherent non-orthogonal MPSK is shown.

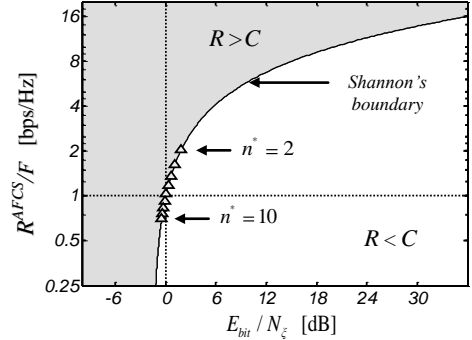

Fig. 5. Empirical assessment of AFCS efficiency at the bandwidth-power efficiency plane.

The results of simulations also confirm this conclusion. The presented in Fig. 5 points "▲" refer to empirical bit-rates $R_{n^*}/F_0$ at the threshold AFCS output computed for corresponding values $E^{bit}/N_\xi$. Values $R_{n^*}/F_0$ were computed using (44),(49) under parameters: $A = 5$mV; $F = 2.5$ kHz; $\sigma_0^2 = 62,5$ mW; $N_\xi = 10^{-10}$ Wt/Hz; $\alpha = 4$, ($\mu \sim 10^{-5}$); $\sigma_v^2 = \sigma_0^2(1+Q^2)^{-n^*}$ mW. Values $E^{bit}/N_\xi$ were computed using the right side of (61) under $F_0 = n^* F$. The experiments were carried out consequently for $n^* = 2,...,10$.

The gain in bit-rate of threshold AFCS ($n = n^*$) in comparison with optimal CS with the same but not adjusted modulator M1 (no feedback) can be assessed by the formula:

$$\rho = \left.\frac{R_n}{R_n^{CS}}\right|_{n=n^*} = \left.\frac{E_{CS}^{bit}}{E^{bit}}\right|_{n=n^*} = \frac{F_0}{F}\frac{\log_2\left(1+\frac{F_0}{F}Q_{CS}^2\right)}{\log_2\left(1+Q_{CS}^2\right)} \quad (62)$$

where $Q_{CS}^2 = W^{sign}/N_\xi F$ is SNR at the input of linear receiver of optimal CS without feedback (with non-adaptive modulator M1, see Sect. II), and bit-rate $R_n^{CS}$ is computed using (31),(32),(44). As illustration, in Tab. 1, the bit-rate gains computed for $n^* = 1,...,5$ according to (62) under parameters used in simulations (previous point) are presented.

Tab.1 Bit-rate gain versus threshold number of cycles

| $n^*$ | 1 | 2 | 3 | 4 | 5 |
|---|---|---|---|---|---|
| $\rho$ | 1 | 2.6 | 4.4 | 6.4 | 8.6 |

Summary: dependences (58)-(61) are obtained directly from the algorithm (24)-(27) that confirms the threshold AFCS work directly at the Shannon's boundary. One should emphasize that this result is valid only for the threshold systems and is not valid, if optimal AFCS is used in non-threshold mode.

*C. Other factors influencing the efficiency of AFCS*

1. Main factor limiting the range of AFCS application is a time delay necessary for computing and delivering the adjusting controls to TU and preparation of system for the next cycle of transmission. Maximal distance $r^{max}$ of AFCS reliable work can be found from the condition $\Delta t_0 > \Delta t_{proc} + 2r/c$, where $\Delta t_{proc}$ describes the loss of time in computing and receiving units of the system, $\Delta t_0 = 1/2F_0$, $c = 3 \cdot 10^{-8} m/s$.

The latter inequality gives the following evaluation:

$$r^{\max} \leq \frac{1}{2}c(\Delta t_0 - \Delta t_{proc}) \quad (63)$$

Formula (63) can be interpreted as the trade-off between the bandwidth of transmitted signals and distance of reliable signal transmission. One should notice that limitation (63) is not crucial and can be removed in more advanced versions of the systems.

2. According to (23), changes of distance $r$ between the TU and BS, as well as of channel gain $\gamma$ change the SNR at the channel Ch1 output. For this reason, AFCS optimal under SNR $Q^2$ at definite "reference" distance will be working non-optimally at the other distances. In wireless sensor nets or other applications with the fixed location of TU, this difficulty can be solved by direct measurement of distances between TU and BS (or of the full gains $\gamma/r$). In mobile applications, maintaining of threshold mode of transmission requires development of additional procedures enabling current assessment of the channel Ch1 gain.

3. For poor feedback channels with real MSE $\bar{\sigma}_v^2$ of feedback transmission greater than the nominal MMSE of threshold AFCS ($\bar{\sigma}_v^2 > \sigma_v^2 = \sigma_{out}^2$) the threshold interval will be shortened, and efficiency of AFCS will be substantially lower than nominal. However, lack of special constraints on the power of signals transmitted by BS allows to decrease $\bar{\sigma}_v^2$ up to the values $\bar{\sigma}_v^2 \leq \sigma_v^2$ at the distances $r \leq r^{\max}$ and to preserve the required quality of the signal transmission.

These and other questions concerning practical applications of results of the paper require independent consideration.

## VI. CONCLUDING REMARKS

The results of research show that concurrent optimization of the transmitting and receiving parts of AFCS, as well as its application in threshold mode make AFCS ideal communication system which transmit the signals with a bit-rate equal to the capacity of the forward channel under given BER $\mu$.

The key to accurate solution of this task was developed by author original approach [3],[4] to full optimization of the analog-digital estimation systems with adaptively adjusted analogue part. The key element of AFCS architecture, which enabled application of approach [3],[4] was the feedback channel. Just its optimal utilization has permitted to impart simple PAM transmitter the properties of ideal coding unit. However, in communication systems without feedback, analogue transmission loses any advantages in comparison with digital transmission. Else one key was statistical fitting condition (7),(8) which was not used by other authors and enabled accurate solution of full optimization tasks both in the mentioned works and in present paper.

The systems considered in the paper have indisputable advantage: their transmitting units do not contain analog-digital converters and coding units. That means they may find sufficiently wide own field of applications in designing the high-efficient extremely low energy/size/cost transmission units for short and middle range communications (wireless sensor nets, RFID, Bluetooth-type units, etc.).

The obtained results enable interesting theoretically and useful for practice development and generalization.

REFERENCES

[1] S. Haykin, *Communication Systems*, NY, J. Wiley, 2001.
[2] S. Verdu, Fifty years of Shannon theory, *IEEE Trans. Inf. Theory*, vol. 44, no. 6, 1998, pp. 1998 2057.
[3] A. Platonov, "Optimal identification of regression-type processes under adaptively controlled observations", IEEE Trans. on Sign. Proc., vol. 42, no. 9, Sept. 1994, pp. 2280-2291.
[4] A. Platonov, *Analytical methods of analog-digital adaptive estimation systems design,* Warsaw, Publishing House Warsaw Univ. of Techn., series "Electronics", vol. 154, 2006 (in Polish).
[5] A. Platonov, "Optimization of adaptive communication systems with feedback channels, *IEEE Wireless Comm. and Net. Conf. WCNC 2009*, Budapest, Apr. 2009, (IEEE Xplore).
[6] C. E. Shannon, A Mathematical Theory of Communication, *The Bell System Technical Journal,* vol. 27, pp. 379–423, 623–656, July, October, 1948.
[7] C. E. Shannon, "Coding theorems for a discrete source with a fidelity criterion," *In IRE Int. Conv. Rec., part 4,* vol. 7, 1959, pp. 142-163, Reprinted with changes in *Information and Decision Processes,* R. E. Machol, NY, McGraw-Hill, 1960, pp. 93-126.
[8] T. Kailath, "An Application of Shannon's Rate-Distortion Theory to Analog Communication over Feedback Channels", *Proc. IEEE,* vol. 55, no. 6, 1967, pp. 1102-1103.
[9] T.J. Goblick, "Theoretical limitations on the transmission of data from analog sources", *IEEE Trans. on Inform. Theory*, vol. 11, no.4, 1965, pp. 558 – 567.
[10] J. K. Omura, "Signal optimization for channels with feedback", PhD dissertation, Rept. SEL-6-068, Stanford, US, Aug. 1966.
[11] J. K. Omura. "Optimum linear transmission of analog data for channels with feedback". *IEEE Trans. on Inform. Theory,* vol.14, no. 1, 1968, pp. 38–43.
[12] J.P.M. Schalkwijk, L.I. Bluestein, "Transmission of Analog Waveforms through Channels with Feedback", *IEEE Trans. on Inform. Theory,* vol. 13, no.4, 1967, pp.617-619.
[13] R.G. Gallager. *Information Theory and Reliable Communication.* John Wiley and Sons, 1968.
[14] E. Baccarelli, R. Cusani, "Linear feedback communication systems with Markov sources: optimal system design and performance evaluation, *IEEE Trans. on Inf. Theory,* vol. 41, no. 6, 1995, pp.1868-1876.
[15] M. Gastpar, "To code or not to code", DSc dissertation, Dept. Inform. and Comm., EPFL, Lausanne, 2002.
[16] A.V. Balakrishnan, *Kalman filtering theory, Optimization Software,* Inc., Publications Division, New York, 1984.
[17] H.L Van Trees., *Detection, estimation and modulation theory*, J. Wiley, New York, 1972.

## Appendix A

Let us assume that over-modulation has occurred at the $k$-th cycle of transmission. Then, according to (4), corrupted observation $\tilde{y}_k$ at the unit DM1 output takes the form:

$$\tilde{y}_k = A + \xi_k \qquad (A.1)$$

(the sign of $A$ depends on the upper or lower saturation level has been crossed but it does not influence the further results). In this case, conditional PDF of the corrupted observation $p_d(\tilde{y}_k \mid \tilde{y}_1^{k-1}, x)$ does not depend on the conditions (index $d$ denotes the distorted values):

$$p_d(\tilde{y}_k \mid \tilde{y}_1^{k-1}, x_k) = p_d(\tilde{y}_k) = \frac{1}{\sqrt{2\pi\sigma_\xi^2}} \exp\left[-\frac{(\tilde{y}_k - A)^2}{2\sigma_\xi^2}\right] \qquad (A.2)$$

and following relationships for PDFs are valid:

$$p_d(\tilde{y}_k \mid \tilde{y}_1^{k-1}) = p_d(\tilde{y}_k), \quad p_d(\tilde{y}_1^k) = p_d(\tilde{y}_k) p(\tilde{y}_1^{k-1})$$

$$p_d(\tilde{y}_1^k \mid x_{k+1}) = p_d(\tilde{y}_k \mid \tilde{y}_1^{k-1}, x_{k+1}) p(\tilde{y}_1^{k-1} \mid x_{k+1}) =$$
$$= p_d(\tilde{y}_k) p(\tilde{y}_1^{k-1} \mid x_{k+1}) \qquad (A.3)$$

Appearance of corrupted observation does not change PDF of prediction $p_d(x_{k+1} \mid \tilde{y}_1^k)$. Really, using formulas (A.3), one may obtain:

$$p_d(x_{k+1} \mid \tilde{y}_1^k) = \frac{p_d(\tilde{y}_1^k \mid x_{k+1}) p(x_{k+1})}{p(\tilde{y}_1^k)} =$$

$$= \frac{p_d(\tilde{y}_k) p(\tilde{y}_1^{k-1} \mid x_{k+1}) p(x_{k+1})}{p_d(\tilde{y}_k) p(\tilde{y}_1^k)} = p(x_{k+1} \mid \tilde{y}_1^{k-1}) = \qquad (A.4)$$

$$= \frac{1}{\sqrt{2\pi(\sigma_v^2 + P_{k-1})}} \exp\left[-\frac{(x_{k+1} - \hat{x}_{k-1})^2}{2(\sigma_v^2 + P_{k-1})}\right]$$

Formula (A.4) permits to conclude that optimal estimates $\hat{x}_{k+1,k}^{(d)}, \hat{x}_k^{(d)}$ (if over-modulation can be registered) are equal to the non-corrupted value $\hat{x}_{k-1}$ computed in previous cycle:

$$\hat{x}_{k+1,k}^{(d)} = \hat{x}_k^{(d)} = \hat{x}_{k-1} \qquad (A.5)$$

Formula (A.5) shows that the optimal reaction on over-modulation is a rejection of the corrupted observations. However, algorithm (24) computes the corrupted estimates $\bar{\hat{x}}_k^{(d)}$ continuously that results in they are non-optimal and add erroneous information to the previous estimate:

$$\bar{\hat{x}}_k^{(d)} = \hat{x}_{k-1} + L_k(A + \xi_k). \qquad (A.6)$$

In effect, MMSE of the estimates $\bar{\hat{x}}_k^{(d)}$ is greater then the previous one:

$$P_k = E[(x - \bar{\hat{x}}_k^{(d)})^2] = P_{k-1} + L_k^2 \sigma_\xi^2 \qquad (A.7)$$

The said above means that in $(k+1)$-th cycle fitting condition (15) will be violated, and probability of restoring the linear mode of transmission in this cycle takes the value:

$$\Pr_{k+1}^{lin} = \int_{B_{k+1} - \frac{1}{M_{k+1}}}^{B_{k+1} + \frac{1}{M_{k+1}}} p_d(x_{k+1} \mid \tilde{y}_1^k)\, dx_k =$$

$$= \frac{1}{\sqrt{2\pi(\sigma_v^2 + P_{k-1})}} \int_{\bar{x}_k - \frac{1}{M_{k+1}}}^{\bar{x}_k + \frac{1}{M_{k+1}}} e^{-\frac{[x - \hat{x}_{k-1}]^2}{2(\sigma_v^2 + P_{k-1})}}\, dx_k =$$

$$= \Phi\left(\frac{M_{k+1}(\bar{x}_k - \hat{x}_{k-1}) + 1}{M_{k+1}\sqrt{\sigma_v^2 + P_{k-1}}}\right) - \Phi\left(\frac{M_{k+1}(\bar{x}_k - \hat{x}_{k-1}) - 1}{M_{k+1}\sqrt{\sigma_v^2 + P_{k-1}}}\right). \qquad (A.8)$$

In (A.8), it is taken into account that signal transmitted to TU is determined by (A.6): $B_k = \bar{\hat{x}}_k^{(d)}$. Parameters $M_{k+1}$ are computed independently according to (19). Denoting the arguments of error functions in (A.8) as $\beta_1; \beta_2$ and taking into account (A.6), (19), (25), one may obtain the relationships:

$$\beta_{1,2} \triangleq \frac{M_{k+1}(\bar{x}_k - \hat{x}_{k-1}) \pm 1}{M_{k+1}\sqrt{\sigma_v^2 + P_{k-1}}} = \frac{1}{\sqrt{\sigma_v^2 + P_{k-1}}} \left[L_k(A + \xi_k) \pm \frac{1}{M_{k+1}}\right] =$$

$$= \frac{1}{M_k \sqrt{\sigma_v^2 + P_{k-1}}} \left[\left(1 - \frac{P_k}{P_{k-1}}\right)\frac{(A + \xi_k)}{A} \pm \frac{M_k}{M_{k+1}}\right] = \qquad (A.9)$$

$$= \alpha\left[\left(1 - \frac{P_k}{P_{k-1}}\right)\left(1 + \frac{\xi_k}{A}\right) \pm \sqrt{\frac{\sigma_v^2 + P_k}{\sigma_v^2 + P_{k-1}}}\right]$$

In Sect. III, it was shown that, in pre-threshold interval $1 \le k < n^*$, $P_{k-1}/\sigma_v^2 \gg 1 + Q^2$ and $P_k = \sigma_0^2(1 + Q^2)^{-k}$. This allows us to write (A.9), in this interval, in the form:

$$\beta_{1,2} = \alpha\left[\left(\frac{Q^2}{1+Q^2}\right)\left(1 + \frac{\xi_k}{A}\right) \pm \frac{1}{\sqrt{1+Q^2}}\right] \underset{Q^2 \gg 1}{=} \alpha\left(1 \pm \frac{1}{Q} + \frac{\xi_k}{A}\right) \qquad (A.10)$$

In this case, probability (A.8) takes the value:

$$\Pr_{k+1}^{lin} = \Phi\left[\alpha\left(1 + \frac{1}{Q}\right) + \frac{\xi_k}{A}\right] - \Phi\left[\alpha\left(1 - \frac{1}{Q}\right) + \frac{\xi_k}{A}\right] \qquad (A.11)$$

Assuming that both $1/Q$ and $\xi_k/A$ have the values always less than unity, (A.11) permits to obtain the assessment:

$$\Pr_{k+1}^{lin} = \Phi\left[\alpha + \alpha\left(\frac{\xi_k}{A} + \frac{1}{Q}\right)\right] - \Phi\left[\alpha + \alpha\left(\frac{\xi_k}{A} - \frac{1}{Q}\right)\right] \approx$$

$$\approx \frac{1}{\sqrt{2\pi}} e^{-\frac{\alpha^2}{2}}\left[\alpha\left(\frac{\xi_k}{A} + \frac{1}{Q}\right) - \alpha\left(\frac{\xi_k}{A} - \frac{1}{Q}\right)\right] = \alpha Q^{-1}\sqrt{\frac{2}{\pi}} e^{-\frac{\alpha^2}{2}} \ll 1 \qquad (A.12)$$

In the "post-threshold" interval $k > n^*$ MSE of estimates satisfies the inequality $P_{k-1}/\sigma_v^2 \ll 1$, and takes value (29). In this case (A.9) permits to obtain valid for $k > n^*$ assessment:

$$\beta_{1,2} = \alpha\left[\left(\frac{1}{k - n^* + 1}\right)\left(1 + \frac{\xi_k}{A}\right) \pm 1\right] \qquad (A.13)$$

Substituting (A.13) into (A.9) and repeating operations used in derivation of (A.12) leads to the relationship:

$$\Pr_{k+1}^{lin} = \Phi\left[\left(\frac{\alpha}{k - n^* + 1} + \frac{\xi_k}{A}\right) + \alpha\right] - \Phi\left[\left(\frac{\alpha}{k - n^* + 1} + \frac{\xi_k}{A}\right) - \alpha\right] =$$

$$= \Phi\left[\alpha + \left(\frac{\alpha}{k - n^* + 1} + \frac{\xi_k}{A}\right)\right] + \Phi\left[\alpha - \left(\frac{\alpha}{k - n^* + 1} + \frac{\xi_k}{A}\right)\right] \approx 2\Phi(\alpha) = 1 - \mu \qquad (A.14)$$

The results of analysis show:

1. Appearance of over-modulation causes appearance of abnormal errors in estimates.

2. Probability of restoration of the linear mode of transmission after appearance of over-modulation in the pre-threshold interval $1 \le k < n^*$ is close to zero. In the post-threshold interval $k > n^*$, it takes the values close to unity $(1 - \mu)$